\documentclass[10pt]{article}
\usepackage{latexsym}
\usepackage{amssymb}
\usepackage{amsmath}
\usepackage{amscd}
\usepackage{amsthm}
\usepackage[left=2cm,top=2.5cm,right=2.5cm,bottom=1.5cm]{geometry}
\usepackage[dvips]{graphicx}
\usepackage{epstopdf}
\usepackage{hyperref}
\begin{document}
\begin{center}
\large{\bf{Two-Fluid Dark Energy Models in Bianchi Type-III Universe with Variable Deceleration Parameter}} \\
\vspace{10mm}
\normalsize{Hassan Amirhashchi$^1$, Anirudh Pradhan$^2$, Rekha Jaiswal$^3$} \\
\vspace{5mm}
\normalsize{$^{1}$Laboratory of Computational Sciences and Mathematical Physics, Institute for
Mathematical Research, University Putra Malaysia, 43400 UPM, Serdang, Selangor D.E., Malaysia\\$^{1}$Department of Physics, Mahshahr Branch, Islamic Azad University, Mahshahr, Iran \\
\vspace{2mm}
$^1$E-mail: hashchi@yahoo.com; h.amirhashchi@mahshahriau.ac.ir} \\
\vspace{5mm}
\normalsize{$^{2,3}$Department of Mathematics, Hindu Post-graduate College, Zamania-232 331, Ghazipur, India \\
\vspace{2mm}
$^2$E-mail: pradhan@iucaa.ernet.in; pradhan.anirudh@gmail.com \\
\vspace{2mm}
$^3$E-mail: rekhajaiswal68@yahoo.com} \\

\end{center}
\vspace{10mm}
\begin{abstract}
Some new exact solutions of Einstein's field equations have come forth within the scope of a spatially homogeneous
and anisotropic Bianchi type-III space-time filled with barotropic fluid and dark energy by considering a variable
deceleration parameter. We consider the case when the dark energy is minimally coupled to the perfect fluid as well
as direct interaction with it. Under the suitable condition, the anisotropic models approach to isotropic scenario.
We also find that during the evolution of the universe, the equation of state (EoS) for dark energy $\omega^{(de)}$,
in both cases, tends to $-1$ (cosmological constant, $\omega^{(de)} = -1$), by displaying various patterns as time
increases, which is consistent with recent observations. The cosmic jerk parameter in our derived models are in good
agreement with the recent data of astrophysical observations under appropriate condition. It is observed that the
universe starts from an asymptotic Einstein static era and reaches to the $\Lambda$CDM model. So from recently developed
Statefinder parameters, the behavior of different stages of the universe has been studied. The physical and geometric
properties of cosmological models are also discussed.
\end{abstract}

\section{Introduction}
There is good evidence that a mysterious form of dark energy (DE) accounts for about two-third of matter and energy
in the Universe. The direct evidence comes from distance measurements of type Ia supernovae (SNe Ia) as standard
candles which indicate the expansion of the universe is speeding up, not slowing down \cite{ref1}$-$\cite{ref3}.
In addition, measurements of cosmic microwave background \cite{ref4} and the galaxy power spectrum \cite{ref5}
also indicate the existence of the dark energy. These observations have reopened the quest for the cosmological
constant which was introduced by Einstein \cite{ref6} in his field equations, but later abandoned \cite{ref7}
and infamously cited as his greatest blunder \cite{ref8}. The cosmological constant can be considered as new kind of
``world matter`` \cite{ref9} and be identified with the energy density of the vacuum \cite{ref10}. The simplest
candidate for dark energy is the energy density of the quantum vacuum (or cosmological constant) for which
$p = -\rho$. However, the inability of particle theorists to complete the energy of the quantum vacuum - contributions
from well understood physics amount to $10^{55}$ times critical density - casts a dark shadow on the cosmological constant
\cite{ref11}. In addition to this, a number of viable models for DE have been fabricated. These scenarios include,
quintessence \cite{ref12,ref13}, modified gravity \cite{ref14}$-$\cite{ref20}, tachyon \cite{ref21}
arising in string theory \cite{ref22}, quintessential inflation \cite{ref23}, chaplygin gas as well as
generalized chaplygin gas \cite{ref24}$-$\cite{ref27}, cosmological nuclear energy \cite{ref28}, equation of
state (EoS) parameter \cite{ref29}$-$\cite{ref39}, braneworld \cite{ref40,ref41} and interacting dark energy
models \cite{ref42}$-$\cite{ref48}. Therefore some form of dark energy whose fractional energy density is
about $\Omega^{(de)} = 0.70$ must exist in the Universe to drive this acceleration. This fact can be put in
agreement with the theory, if one assumes that the Universe is basically filled with so-called dark energy.
Evolution of the equation of state (EoS) of dark energy $\omega^{(de)} = \frac{p^{(de)}}{\rho^{(de)}}$ transfers from
$\omega^{(de)} > -1$ in the near past (quintessence region) to $\omega^{(de)} < -1$ at recent stage (phantom region)
\cite{ref49}$-$\cite{ref51}. So another cosmological coincidence problem may be proposed: why $ \omega^{(de)} = −1$
crossing is occurred at the present time \cite{ref52}. When SNe results are combined with five-year WMAP, it is
found that $-1.38 < \omega^{(de)} < 0.86$ \cite{ref53}$-$\cite{ref55}.  For recent review, the readers are advised
to see the references of Padmanabhan \cite{ref56}, Copeland et al. \cite{ref57}, Perivolaropoulos
\cite{ref58}, Jassal et al. \cite{ref59} and Miao et al. \cite{ref60}. \\

The simplest model of the observed universe is well represented by Friedmann-Robertson-Walker (FRW) models, which
are both spatially homogeneous and isotropic. These models in some sense are good global approximation of the
present-day universe. But it is also believed that in the early universe the FRW model does not give a correct
matter description. The anomalies found in the cosmic microwave background (CMB) and the large structure
observations stimulated a growing interest in anisotropic cosmological models of universe. Observations by the
Differential Microwave Radiometers (DMR) on NASA's Cosmic Background Explorer (COBE) registered anisotropy in various
angle scales. Ellis \cite{ref61} pointed out that although the observed universe seems to be almost isotropic on
large scales, the early and/or very late universe could be anisotropic. Furthermore, the interest in such models
was encouraged in recent years due to the argument that going on the analysis and the interpretation of the WMAP
\cite{ref62}$-$\cite{ref64} data, whether they need a Bianchi type morphology to be explained successfully
\cite{ref65}$-$\cite{ref70}. Thus, in recent years Bianchi universes have been gaining an increasing interest of
observational cosmology. \\

In earlier, Xin \cite{ref71} studied an interacting two-fluid scenario for quintom dark energy. Xin-He et
al. \cite{ref72} considered Friedmann cosmology with a generalized EoS and bulk viscosity to explain DE
dominated universe. Recently, several authors \cite{ref73}$-$\cite{ref85} have examined and discussed the DE
models in different context of use. \\

Incited by above discussions, in this paper, we purport to probe the evolution of the dark energy parameter
within the framework of a Bianchi type-III cosmological model filled with two fluids by considering a variable
deceleration parameter (DP). In doing so we consider both non-interacting and interacting cases. This paper is
organized as follows: the metric and the field equations are presented in Sec. $2$. Section $3$ deals with the
exact solutions of the field equations. In Sec. $4$ we obtain the non-interacting two-fluid model and its
geometric and physical significance are discussed. Section $5$ deals with cosmic jerk parameter. In Sec. $6$
interacting two-fluid scenario is presented. Concluding remarks are given in Sec. $7$.
\section{The Metric and Field Equations}
We consider the Bianchi type-III metric as
\begin{equation}
\label{eq1}
ds^{2} = -dt^{2} + A^{2}(t)dx^{2}+B^{2}(t)e^{-2\alpha x}dy^{2}+C^{2}(t)dz^{2},
\end{equation}
where $A(t), B(t)$ and $C(t)$ are functions of time only. \\

We define the following physical and geometric parameters to be used in formulating the law and further in solving
the Einstein's field equations for the metric (\ref{eq1}). \\

The average scale factor $a$ of Bianchi type-III model (\ref{eq1}) is defined as
\begin{equation}
\label{eq2} a = (ABC)^{\frac{1}{3}}.
\end{equation}
A volume scale factor V is given by
\begin{equation}
\label{eq3} V = a^{3} = ABC.
\end{equation}
We define the generalized mean Hubble's parameter $\rm H$ as
\begin{equation}
\label{eq4} H = \frac{1}{3}(H_{x} + H_{y} + H_{z}),
\end{equation}
where $\rm H_{x} = \frac{\dot{A}}{A}$, $\rm H_{y} = \frac{\dot{B}}{B}$ and $\rm H_{z} = \frac{\dot{C}}{C}$ are the
directional Hubble's parameters in the directions of $x$, $y$ and $z$ respectively. A dot stands for differentiation
with respect to cosmic time $t$. \\

From Eqs. (\ref{eq2})-(\ref{eq4}), we obtain
\begin{equation}
\label{eq5} H = \frac{1}{3}\frac{\dot{V}}{V} = \frac{\dot{a}}{a} = \frac{1}{3}\left(\frac{\dot{A}}{A} +
\frac{\dot{B}}{B} + \frac{\dot{C}}{C}\right).
\end{equation}
The physical quantities of observational interest in cosmology i.e. the expansion scalar $\theta$, the average
anisotropy parameter $Am$ and the shear scalar $\sigma^{2}$ are defined as
\begin{equation}
\label{eq6} \theta = u^{i}_{;i} = \left(\frac{\dot{A}}{A} + \frac{\dot{B}}{B} + \frac{\dot{C}}{C} \right),
\end{equation}
\begin{equation}
\label{eq7} \sigma^{2} =  \frac{1}{2} \sigma_{ij}\sigma^{ij} = \frac{1}{2}\left[\frac{\dot{A}^{2}}{A^{2}} +
\frac{\dot{B}^{2}}{B^{2}} + \frac{\dot{C}^{2}}{C^{2}}\right] - \frac{\theta^{2}}{6},
\end{equation}
\begin{equation}
\label{eq8} A_{m} = \frac{1}{3}\sum_{i = 1}^{3}{\left(\frac{\triangle H_{i}}{H}\right)^{2}},
\end{equation}
where $\triangle H_{i} = H_{i} - H (i = x, y, z)$ represents the directional Hubble parameter in the direction of
$x$, $y$, $z$ respectively. $A_{m} = 0$ corresponds to isotropic expansion. \\

The Einstein's field equations ( in gravitational units $8\pi G = c = 1 $) read as
\begin{equation}
\label{eq9} R^{i}_{j} - \frac{1}{2} R g^{i}_{j} = - T^{(m)i}_{j} - T^{(de)i}_{j},
\end{equation}
where $T^{(m)i}_{j}$ and $T^{(de)i}_{j}$ are the energy momentum tensors of perfect fluid and DE, respectively. These
are given by
\[
  T^{(m)i}_{j} = \mbox{diag}[-\rho^{(m)}, p^{(m)}, p^{(m)}, p^{(m)}],
\]
\begin{equation}
\label{eq10} ~ ~ ~ ~ ~ ~ ~ ~  = \mbox{diag}[-1, \omega^{(m)}, \omega^{(m)}, \omega^{(m)}]\rho^{m},
\end{equation}
and
\[
 T^{(de)i}_{j} = \mbox{diag}[-\rho^{(de)}, p^{(de)}, p^{(de)}, p^{(de)}],
\]
\begin{equation}
\label{eq11} ~ ~ ~ ~ ~ ~ ~ ~ ~ ~ ~ ~ ~ ~ = \mbox{diag}[-1, \omega^{(de)}, \omega^{(de)}, \omega^{(de)}]\rho^{(de)},
\end{equation}
where $\rho^{(m)}$ and $p^{(m)}$ are, respectively the energy density and pressure of the perfect fluid component or
ordinary baryonic matter while $\omega^{(m)} = p^{(m)}/\rho{(m)}$ is its EoS parameter. Similarly, $\rho^{(de)}$ and
$p^{(de)}$ are, respectively the energy density and pressure of the DE component while $\omega^{(de)} = p^{(de)}/\rho^{(de)}$
is the corresponding EoS parameter. We assume the four velocity vector $u^{i} = (1, 0, 0, 0)$ satisfying $u^{i}u_{j} = -1$. \\

In a co-moving coordinate system ($u^{i} = \delta^{i}_{0}$), Einstein's field equations (\ref{eq9}) with (\ref{eq10})
and (\ref{eq11}) for Bianchi type-III metric (\ref{eq1}) subsequently lead to the following system of equations:
\begin{equation}
\label{eq12} \frac{\ddot{B}}{B} + \frac{\ddot{C}}{C} + \frac{\dot{B}\dot{C}}{BC}
= -\omega^{(m)}\rho^{(m)} - \omega^{(de)}\rho^{(de)},
\end{equation}
\begin{equation}
\label{eq13} \frac{\ddot{C}}{C} + \frac{\ddot{A}}{A} + \frac{\dot{C}\dot{A}}{CA}
= -\omega^{(m)}\rho^{(m)} - \omega^{(de)}\rho^{(de)},
\end{equation}
\begin{equation}
\label{eq14} \frac{\ddot{A}}{A} + \frac{\ddot{B}}{B} + \frac{\dot{A} \dot{B}}{AB} - \frac{\alpha^{2}} {A^{2}}
= -\omega^{(m)}\rho^{(m)} - \omega^{(de)}\rho^{(de)},
\end{equation}
\begin{equation}
\label{eq15} \frac{\dot{A}\dot{B}}{AB} + \frac{\dot{A}\dot{C}}{AC} + \frac{\dot{B}\dot{C}}{BC} - \frac{\alpha^{2}}{A^{2}}
= \rho^{(m)} + \rho^{(de)},
\end{equation}
\begin{equation}
\label{eq16} \alpha\left(\frac{\dot{A}}{A} - \frac{\dot{B}}{B}\right)  = 0.
\end{equation}
The law of energy-conservation equation ($T^{ij}_{;j} = 0$) yields
\begin{equation}
\label{eq17} \dot{\rho}^{(m)} + 3(1 + \omega^{(m)})\rho^{(m)}H + \dot{\rho}^{(de)} +3(1 + \omega^{(de)})\rho^{(de)}H = 0.
\end{equation}
The Raychaudhuri equation is found to be
\begin{equation}
\label{eq18} \dot{\theta} = - \left(1 + 3\omega^{(de)}\right)\rho^{(de)} - \frac{1}{3}\theta^{2} - 2\sigma^{2}.
\end{equation}
\section{Solutions of the Field Equations}
The field equations (\ref{eq12})-(\ref{eq16}) are a system of five linearly independent equations with seven unknown
parameters $A$, $B$, $C$, $\rho^{(m)}$, $p^{(de)}$, $\rho^{(de)}$, $\omega^{(de)}$. Two additional constraints relating
these parameters are required to obtain explicit solutions of the system.\\

Eq. (\ref{eq16}), obviously leads to
\begin{equation}
\label{eq19} B = \ell_{0} A,
\end{equation}
where $\ell_{0}$ is an integrating constant.  \\

Firstly, we assume that the scalar expansion $\theta$ in the model is proportional to the shear scalar. This condition
leads to
\begin{equation}
\label{eq20} A = C^{n},
\end{equation}
where $n$ is a constant.The motive behind assuming this condition is explained with
reference to Thorne \cite{ref86}, the observations of the velocity-red-shift relation for extragalactic
sources suggest that Hubble expansion of the universe is isotropic today within $\approx 30$ per cent
\cite{ref87,ref88}. To put more precisely, red-shift studies place the limit $\frac{\sigma}{H} \leq 0.3$
on the ratio of shear $\sigma$ to Hubble constant $H$ in the neighbourhood of our Galaxy today. Collins
et al. \cite{ref89} have pointed out that for spatially homogeneous metric, the normal congruence to the
homogeneous expansion satisfies that the condition $\frac{\sigma}{\theta}$ is constant. \\

We define the deceleration parameter q as
\begin{equation}
\label{eq21} q \equiv -\frac{\ddot{a}}{a}\left(\frac{\dot{a}}{c}\right)^{-2} = -\frac{a\ddot{a}}{\dot{a}^{2}} =
- \left(\frac{\dot{H} + H^{2}}{H^{2}}\right) = b(t) ~ ~\mbox{say},
\end{equation}
where $a$ is the average scale factor of the universe defined by Eq. (\ref{eq2}) and the dots indicate derivatives by
proper time. The expansion of the universe is said to be ``accelerating'' if $\ddot{a}$ is positive (recent measurements
suggest it is), and in this case the DP will be negative. The minus sign and the name ``deceleration parameter'' are
historical; at the time of definition $q$ was thought to be positive, now it is believed to be negative. Recent observations
\cite{ref1}$-$\cite{ref3} have suggested that the rate of expansion of the universe is
currently accelerating, perhaps due to dark energy. This yields negative values of the DP. \\

The motivation to choose such time dependent DP is behind the fact that the universe is accelerated expansion at present as
observed in recent observations of Type Ia supernova \cite{ref1}$-$\cite{ref3} and CMB anisotropies \cite{ref66,ref90,ref91}
and decelerated expansion in the past. Also, the transition redshift from deceleration expansion to accelerated expansion is
about $0.5$. Now for a Universe which was decelerating in past and accelerating at the present time, the DP must show
signature flipping \cite{ref92}$-$\cite{ref94}. So, there is no scope for a constant DP at the present epoch. So, in general,
the DP is not a constant but time variable. \\

The equation (\ref{eq21}) may be rewritten as
\begin{equation}
\label{eq22} \frac{\ddot{a}}{a} + b\frac{\dot{a}^{2}}{a^{2}} = 0.
\end{equation}
In order to solve the Eq. (\ref{eq22}), we assume $b = b(a)$. It is important to note here that one can assume
$b = b(t) = b(a(t))$, as $a$ is also a time dependent function. It can be done only if there is a one to one
correspondences between $t$ and $a$. But this is only possible when one avoid singularity
like big bang or big rip because both $t$ and $a$ are increasing function.  \\

The general solution of Eq. (\ref{eq22}) with assumption $b = b(a)$, is given by
\begin{equation}
	\label{eq23} \int {e^{\int{\frac{b}{a}}\,da}}\, da = t + k,
\end{equation}
where $k$ is an integrating constant. \\

One can not solve Eq. (\ref{eq23}) in general as $b$ is variable. So, in order to solve the problem completely,
we have to choose $\int{\frac{b}{a}}da$ in such a manner that Eq. (\ref{eq23}) be integrable without any loss of
generality. Hence we consider
\begin{equation}
\label{eq24}\int{\frac{b}{a}}\,da = \ln {L(a)}.
\end{equation}
which does not affect the nature of generality of solution. Hence from Eqs. (\ref{eq23})
and (\ref{eq24}), we obtain
\begin{equation}
\label{eq25}\int {L(a)}\, da = t + k.
\end{equation}
Of course the choice of $L(a)$, in Eq. (\ref{eq25}), is quite arbitrary but, since we are
looking for physically viable models of the universe consistent with observations, we consider
\begin{equation}
\label{eq26}L(a)=\frac{1}{\alpha\sqrt{1 + a^{2}}},
\end{equation}
where $\alpha$ is an arbitrary constant. In this case, on integrating, Eq. (\ref{eq25}) gives the exact solution
\begin{equation}
\label{eq27} a(t) = \sinh{(\beta T)},
\end{equation}
where $T = t + t_{0}$, $t_{0}$ and $\beta$ being constants of integration. We also note that $t + 0$ and $T = \infty$
respectively correspond to the proper time $t = -t_{0}$  and $t = \infty$. The relation (\ref{eq27}) is recently used
by Pradhan et al. \cite{ref38} and Amirhashchi et al. \cite{ref47} in studying dark energy models in Bianchi type-$VI_{0}$
and FRW space-time respectively.  \\

Now, by using (\ref{eq5}), (\ref{eq19}), (\ref{eq20}) and (\ref{eq23}) we can find the metric components as
\begin{equation}
\label{eq28} A = \ell_{1} \sinh^{\frac{3n}{2n + 1}}(\beta T),
\end{equation}
\begin{equation}
\label{eq29} B = \ell_{2} \sinh^{\frac{3n}{2n + 1}}(\beta T),
\end{equation}
\begin{equation}
\label{eq30} C = \ell_{3} \sinh^{\frac{3}{2n + 1}}(\beta T),
\end{equation}
where $\ell_{1} = K^{-\frac{3n}{(2n + 1)}}$, $\ell_{2} = \ell_{0}\ell_{1}$, $\ell_{3} = \ell_{1}^{\frac{1}{n}}$ and
$K$ is an integrating constant. \\

Therefore, the metric (\ref{eq1}) reduces to
\[
ds^{2} = -dt^{2} + \ell_{1}^{2} \sinh^{\frac{6n}{2n+1}}(\beta T) dx^{2} + \ell_{2}^{2} \sinh^{\frac{6n}{2n + 1}}(\beta T)
e^{-2\alpha x}dy^{2}
\]
\begin{equation}
\label{eq31}
\ +\ell_{3}^{2} \sinh^{\frac{6}{2n+1}}(\beta T)dz^{2}.
\end{equation}
In the following sections we deal with two cases, (i) non-interacting two-fluid model and (ii) interacting two-
fluid model.

\section{Non-Interacting Two-Fluid Model}
In this section we assume that two-fluid do not interact with each other. Therefor, the general form of conservation
equation (\ref{eq17}) leads us to write the conservation equation for the barotropic and dark fluid separately as,
\begin{equation}
\label{eq32}\dot{\rho}^{(m)} + 3\frac{\dot{a}}{a}\left(\rho^{(m)} + p^{(m)}\right) =
\dot{\rho}^{(m)} + (1 + \omega^{(m)})\rho^{(m)}(2n + 1)\frac{\dot{C}}{C} = 0,
\end{equation}
and
\begin{equation}
\label{eq33}\dot{\rho}^{(de)} + 3\frac{\dot{a}}{a}\left(\rho^{(de)} + p^{(de)}\right) =
\dot{\rho}^{(de)} + (1 + \omega^{(de)})\rho^{(de)}(2n + 1)\frac{\dot{C}}{C} = 0.
\end{equation}
Integration of (\ref{eq32}) leads to
\begin{equation}
\label{eq34}\rho^{(m)} = \rho_{0}C^{-(2n + 1)(1 + \omega^{(m)})} = \rho_{0}l_{0}\sinh^{-3(1 + \omega^{(m)})}(\beta T),
\end{equation}
where $\rho_{0}$ is an integrating constant and $l_{0} = \ell_{3}^{-(2n + 1)(1 + \omega^{(m)})}$. \\

\begin{figure}[ht]
\begin{minipage}[b]{0.5\linewidth}
\centering
\includegraphics[width=\textwidth]{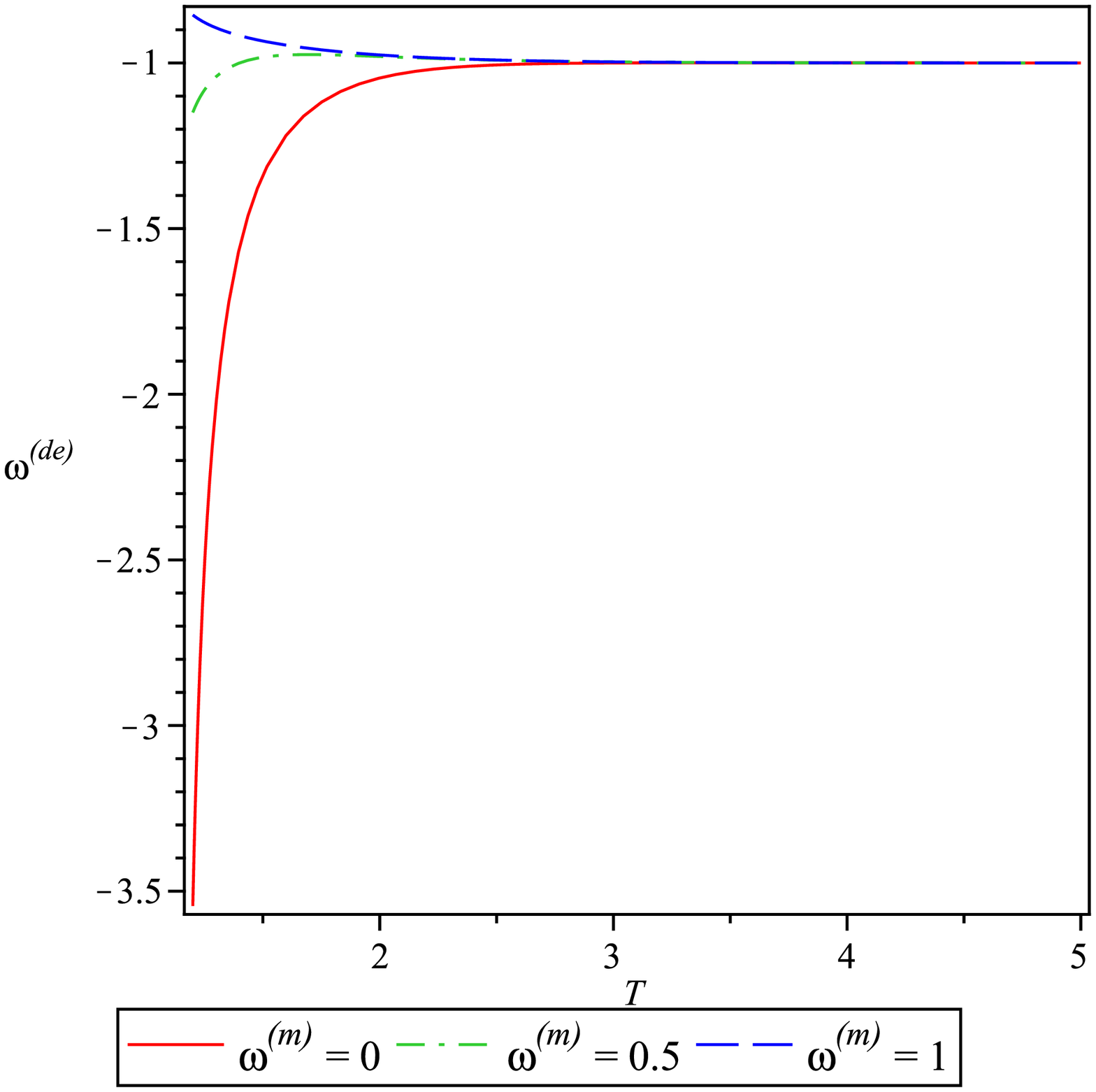}
\caption{The EoS DE parameter $\omega^{(de)}$ versus $T$ for $n = \beta = \alpha = \ell_{3} = l_{0} = 1$,
$\rho_{0} = 10$.}
\label{fig:figure1}
\end{minipage}
\hspace{0.5cm}
\begin{minipage}[b]{0.5\linewidth}
\centering
\includegraphics[width=\textwidth]{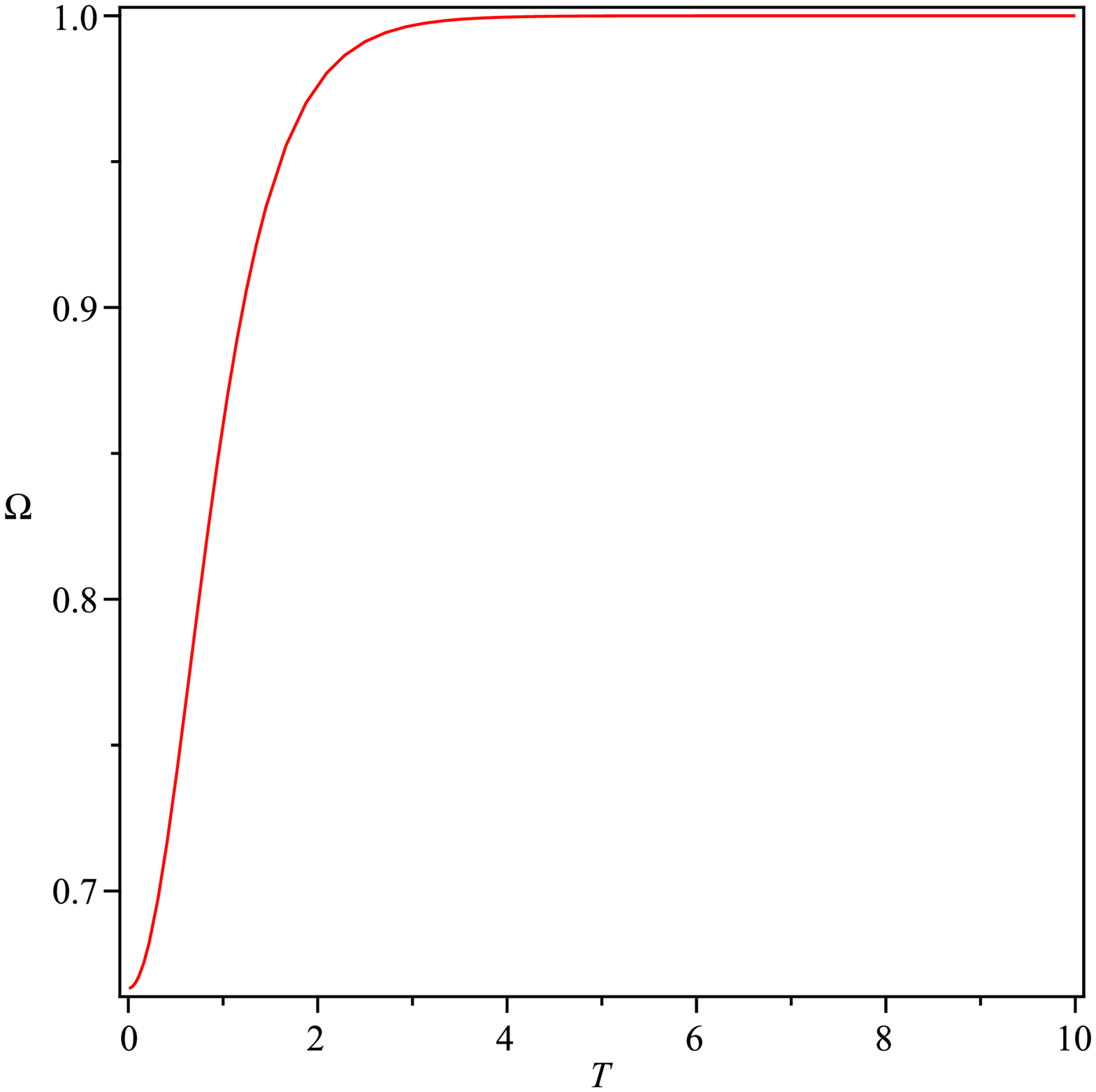}
\caption{The total energy density parameter $\Omega$ vs. $T$ for $n = \beta = \alpha = \ell_{3} = 1$}
\label{fig:figure2}
\end{minipage}
\end{figure}
By using Eqs. (\ref{eq19}), (\ref{eq20}) and (\ref{eq34}) in Eqs. (\ref{eq15}) and (\ref{eq12}), we obtain
\begin{equation}
\label{eq35} \rho^{(de)} = n(n + 2)\frac{\dot{C}^{2}}{C^{2}} - \frac{\alpha^{2}}{C^{2n}} -
\rho_{0}l_{0}\sinh^{-3(1 + \omega^{(m)})}{(\beta T)},
\end{equation}
and
\begin{equation}
\label{eq36} p^{(de)} = - \left[2n\frac{\ddot{C}}{C} + n(3n-2)\frac{\dot{C}^{2}}{C^{2}} - \frac{\alpha^{2}}{C^{2n}}\right]
- \omega^{(m)}\rho_{0}l_{0}\sinh^{-3(1 + \omega^{(m)})}(\beta T).
\end{equation}
Using Eq. (\ref{eq30}) in Eqs. (\ref{eq35}) and (\ref{eq36}), we obtain the values of $\rho^{(de)}$ and $p^{(de)}$ as
\begin{equation}
\label{eq37}
\rho^{(de)} = \frac{9n(n + 2)\beta^{2}}{(2n + 1)^{2}}\coth^{2}(\beta T) - \alpha^{2}\ell^{-2n}_{3}\sinh^{-\frac{6n}
{(2n + 1)}}(\beta T) - \rho_{0}l_{0}\sinh^{-3(1 + \omega^{(m)})}{(\beta T)},
\end{equation}
\begin{equation}
\label{eq38}
p^{de} = -\left[\frac{9(n^{2} + n + 1)}{(2n + 1)^{2}}\coth^{2}(\beta T)- \frac{3(n + 1)\beta^{2}}{(2n + 1)}
cosech^{2}{(\beta T)}\right] - \omega^{(m)}\rho_{0}l_{0}\sinh^{-3(1 +
\omega^{(m)})}(\beta T).
\end{equation}
respectively. \\

The EoS ($\omega^{(de)}$) of DE, for model (\ref{eq31}) is found to be dark energy in term of time as
\begin{equation}
\label{eq39}\omega^{(de)} = - \left[\frac{\frac{9(n^{2} + n + 1)}{(2n + 1)^{2}}\coth^{2}(\beta T)- \frac{3(n + 1)
\beta^{2}}{(2n + 1)}cosech^{2}{(\beta T)} + \omega^{(m)}\rho_{0}l_{0}\sinh^{-3(1 - \omega^{(m)})}(\beta T)}
{\frac{9n(n + 2)\beta^{2}}{(2n + 1)^{2}}\coth^{2}(\beta T) - \alpha^{2}\ell^{-2n}_{3}
\sinh^{-\frac{6n}{(2n + 1)}}{(\beta T)} - \rho_{0}l_{0}\sinh^{-3(1 + \omega^{(m)})}(\beta T)}\right].
\end{equation}
The behaviour of EoS for dark energy in term of cosmic time $T$ is shown in Fig. $1$. We observe from the
Fig. $1$ that for $\omega^{(m)} \geq 1$,  $\omega^{(de)}$ varies from non-dark region crossing quintessence
region and ultimately approaching to cosmological constant region ($\omega^{(de)} = -1$). But for $\omega^{(m)} < 1$,
the variation of $\omega^{(de)}$ starts from from super phantom region ($\omega^{m}\geq 1$, $\omega^{(de)} \leq -3$)
crossing phantom region ($\omega^{(de)} < -1$) and finally approaches to cosmological constant region ($\omega^{(de)} = -1$).
The EoS parameter of the DE may begin in non-dark region or quintessence ($\omega^{de} > -1$) region and tends to $-1$
(cosmological constant, $\omega^{de} = -1$) by exhibiting various patterns as $T$ increases; see Fig. $1$.  On the other
hand, while the current cosmological data from SNIa (Supernova Legacy Survey, Gold Sample of Hubble Space Telescope)
\cite{ref95,ref96}. CMB (WMAP, BOOMERANG) \cite{ref54,ref97} and large scale structure (SDSS) \cite{ref68} data rule
out that $\omega^{(de)} \ll -1$, they mildly favour dynamically evolving DE crossing the PDL (see  [98$-$100, 57]) for
theoretical and observational status of crossing the PDL). Thus, our DE model is in good agreement with well established
theoretical result as well as the recent observations. \\

\begin{figure}[ht]
\begin{minipage}[b]{0.5\linewidth}
\centering
\includegraphics[width=\textwidth]{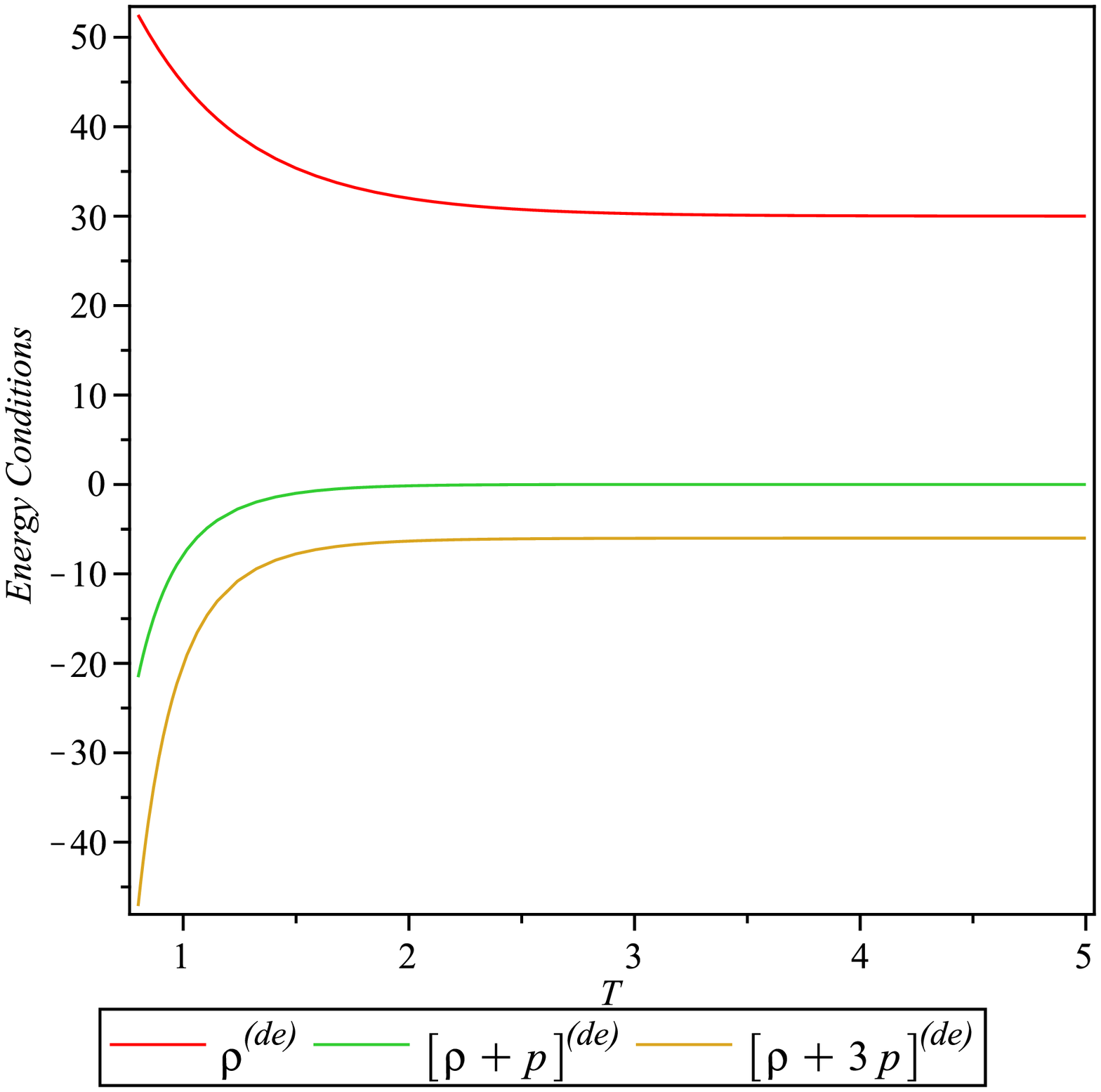}
\caption{The plot of energy conditions versus $T$ for $n = \beta = \alpha = \ell_{3} = l_{0} = 1$, $\rho_{0} = 10,
\omega_{m} = 0.5$}
\label{fig:figure3}
\end{minipage}
\hspace{0.5cm}
\begin{minipage}[b]{0.5\linewidth}
\centering
\includegraphics[width=\textwidth]{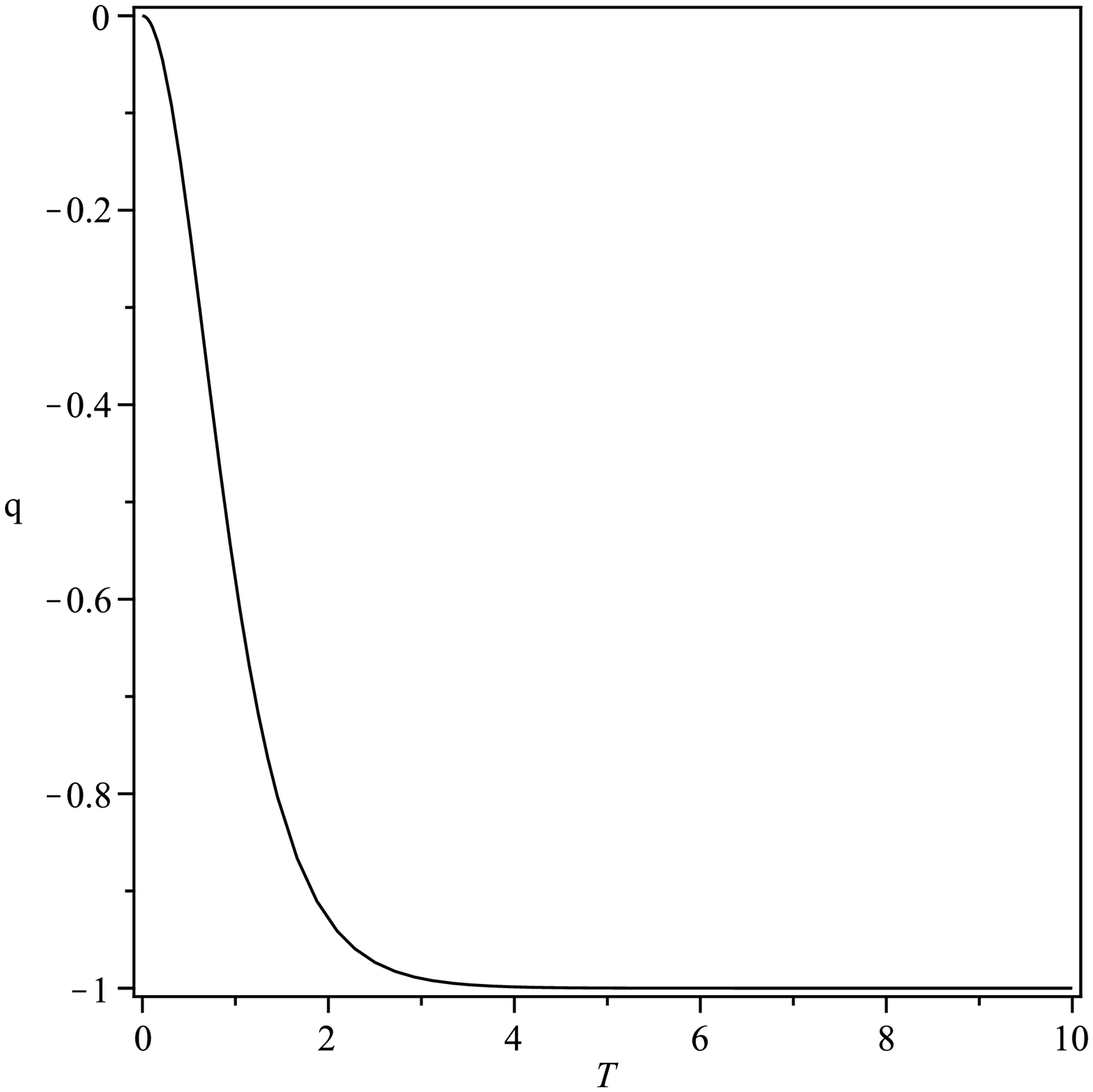}
\caption{The plot of deceleration parameter $q$ Vs. $T$ for $\alpha = 1$}
\label{fig:figure4}
\end{minipage}
\end{figure}
The expressions for the matter-energy density $\Omega^{(m)}$ and dark-energy density $\Omega^{(de)}$ are given by
\begin{equation}
\label{eq40}\Omega^{(m)} = \frac{\rho^{(m)}}{3H^{2}} = \frac{\rho_{0}l_{0}\sinh^{-3(1 - \omega^{(m)})}{(\beta T)}}
{3\beta^{2}\coth^{2}{(\beta T)}},
\end{equation}
and
\begin{equation}
\label{eq41} \Omega^{(de)} =  \frac{\rho^{(de)}}{3H^{2}} = \frac{\frac{9n(n + 2)\beta^{2}}{(2n + 1)^{2}}\coth^{2}(\beta T)
- \alpha^{2}\ell^{-2n}_{3}\sinh^{-\frac{6n}{(2n + 1)}}(\beta T) - \rho_{0}l_{0}\sinh^{-3(1 + \omega^{(m)})}{(\beta T)}}
{3\beta^{2}\coth^{2}{(\beta T)}}.
\end{equation}
respectively. Adding Eqs. (\ref{eq40}) and (\ref{eq41}), we obtain total energy ($\Omega$)
\begin{equation}
\label{eq42}\Omega = \Omega^{(m)} + \Omega^{(de)} = \frac{\frac{9n(n + 2)\beta^{2}}{(2n + 1)^{2}}\coth^{2}(\beta T)
- \alpha^{2}\ell_{3}^{-2n}\sinh^{-\frac{6n}{2n + 1}}{(\beta T)}}{3\beta^{2}\coth^{2}{(\beta T)}}.
\end{equation}
Figure $2$ depicts the variation of the density parameter ($\Omega$) versus cosmic time $T$ during the evolution of
the universe. From the Figure $2$, it can be seen that the total energy density $\Omega$ tends to $1$ for sufficiently
large time which is supported by the current observations. \\

The dark energy with $\omega^{(de)} < -1$, the phantom component of the universe, leads to uncommon cosmological scenarios
as it was pointed out by Caldwell et al. \cite{ref101}. First of all, there is a violation of the dominant energy condition
(DEC), since $\rho^{(de)} + p^{(de)} < 0$. The energy density grows up to infinity in a finite time, which leads to a big rip,
characterized by a scale factor blowing up in this finite time. These sudden future singularities are, nevertheless,
not necessarily produced by a fluid violating DEC. Cosmological solutions for phantom matter which violates the weak
energy condition were found by Dabrowski et al. \cite{ref102}. Caldwell \cite{ref103}, Srivastava \cite{ref24},
Yadav \cite{ref104} have investigated phantom models with $\omega^{(de)} < -1$ and also suggested that at late time, phantom
energy has appeared as a potential DE candidate which violets the weak as well as strong energy condition. \\

The left hand side of energy conditions have been depicted in Figure $3$ for different values of $T$.
From Figure $3$, for $\omega^{(m)}\leq 0.5$ (i.e. phantom model) (also see Figure $1$) , we observe that
$$
(i) ~ ~ \rho^{(de)} \geq 0,  ~ ~ ~ (ii) ~ ~  \rho^{(de)} + p^{(de)} \leq 0,  ~ ~ ~   (iii) ~ ~ \rho^{(de)} + 3p^{(de)} < 0.
$$
Thus, from above expressions, we observe the phantom model violates both the strong and weak energy conditions,
as expected.\\

The expressions for physical parameters such as directional Hubble parameters ($H_{x}$, $H_{y}$, $H_{z}$), the Hubble
parameter ($H$), scalar of expansion ($\theta$), shear scalar ($\sigma$), spatial volume $V$, the anisotropy parameter
($A_{m}$) and the deceleration parameter ($q$) are, respectively, given by
\begin{equation}
\label{eq43} H_{x} = H_{y} = \beta\left(\frac{3n}{2n + 1}\right)\coth{(\beta T)},
\end{equation}
\begin{equation}
\label{eq44}  H_{z} = \beta\left(\frac{3}{2n + 1}\right)\coth{(\beta T)},
\end{equation}
\begin{equation}
\label{eq45} \theta = 3H = 3\beta \coth{(\beta T)},
\end{equation}
\begin{equation}
\label{eq46} \sigma^{2} = 3\left(\frac{n -1}{n + 1}\right)^{2}\beta^{2}\coth^{2}{(\beta T)},
\end{equation}
\begin{equation}
\label{eq47`} V = \sin^{3}{(\beta T)}e^{-\alpha x},
\end{equation}
\begin{equation}
\label{eq48} A_{m} = 2\left(\frac{n - 1}{n + 1}\right)^{2}.
\end{equation}
\begin{equation}
\label{eq49}q = -\tanh^{2}{(\beta T)}.
\end{equation}
It is observed that at $T = 0$, the spatial volume vanishes and other parameters $\theta$, $\sigma$, $H$ diverge.
Hence the model starts with a big bang singularity at $T = 0$. This is a Point Type singularity \cite{ref105}
since directional scale factor $A(t)$, $B(t)$ and $C(t)$ vanish at initial time. For $n = 1$, $A_{m} = 0$. Hence
the model is isotropic when $n = 1$. Fig. $4$ shows the evolution trajectory of deceleration parameter, from which
we observe that the evolution of the universe is in accelerating phase at present era. This is good agreement with
recent observations that our universe is in an accelerated expansion moment \cite{ref1}$-$\cite{ref3}. \\

We also solve for the deceleration parameter $q$ as a function of the redshift $z = -1 + \frac{a_{0}}{a}$,
where $a_{0}$  is the present value of the scale factor (i.e at $z = 0$). It is given by
\begin{equation}
\label{eq50}q(z) = -\Biggl[\tanh \left\{\sinh ^{-1}{\left(\sqrt{\frac{-q_{0}}{(q_{0} + 1)(z + 1)^{2}}}\right)}.
\right\}\Biggr]^{2},
\end{equation}
Here $q_{0}$ is the present value of the deceleration parameter i.e. at $z = 0$. The analyses of cosmological observations
in literature \cite{ref106}$-$\cite{ref111} furnish the transition redshift of the accelerating expansion as given by
$0.3 < z_{t} < 0.8$.
\section{Cosmic Jerk Parameter}
A convenient method to describe models close to $\Lambda$ CDM is based on the cosmic jerk parameter $j$, a
dimensionless third derivative of the scale factor with respect to the cosmic time \cite{ref112}$-$\cite{ref116}.
A deceleration-to-acceleration transition occurs for models with a positive value of $j_{0}$ and negative
$q_{0}$. Flat $\Lambda$ CDM models have a constant jerk $j = 1$. The jerk parameter in cosmology is defined
as the dimensionless third derivative of the scale factor with respect to cosmic time
\begin{equation}
\label{eq51} j(t) = \frac{1}{H^{3}}\frac{\dot{\ddot{a}}}{a}.
\end{equation}
where the `dots' denote derivatives with respect to cosmic time.
The jerk parameter appears in the fourth term of a Taylor expansion of the scale factor around $a_{0}$
\begin{equation}
\label{eq52} \frac{a(t)}{a_{0}} = 1 + H_{0}(t-t_{0}) - \frac{1}{2}q_{0}H_{0}^{2}(t-t_{0})^{2} +
\frac{1}{6}j_{0}H_{0}^{3}(t-t_{0})^{3} + O\left[(t-t_{0})^{4}\right],
\end{equation}
where the subscript $0$ shows the present value. One can rewrite Eq. (\ref{eq51}) as
\begin{equation}
\label{eq53} j(t) = q + 2q^{2} - \frac{\dot{q}}{H}.
\end{equation}
Using Eq. (\ref{eq50}) in Eq. (\ref{eq53}) we obtain
\begin{equation}
\label{eq54} j(t) = \frac{3(\cosh^{2}{(\beta T)} - 1)}{\cosh^{2}{(\beta T)}}.
\end{equation}
This value overlaps with the value $j\simeq2.16$ obtained from the combination of three kinematical data sets: the
gold sample of type Ia supernovae \cite{ref95}, the SNIa data from the SNLS project \cite{ref96}, and the X-ray
galaxy cluster distance measurements \cite{ref117} for
\begin{equation}
\label{eq55} T = \frac{\cosh{(\sqrt{3})}}{\beta}.
\end{equation}

\begin{figure}[ht]
\begin{minipage}[b]{0.5\linewidth}
\centering
\includegraphics[width=\textwidth]{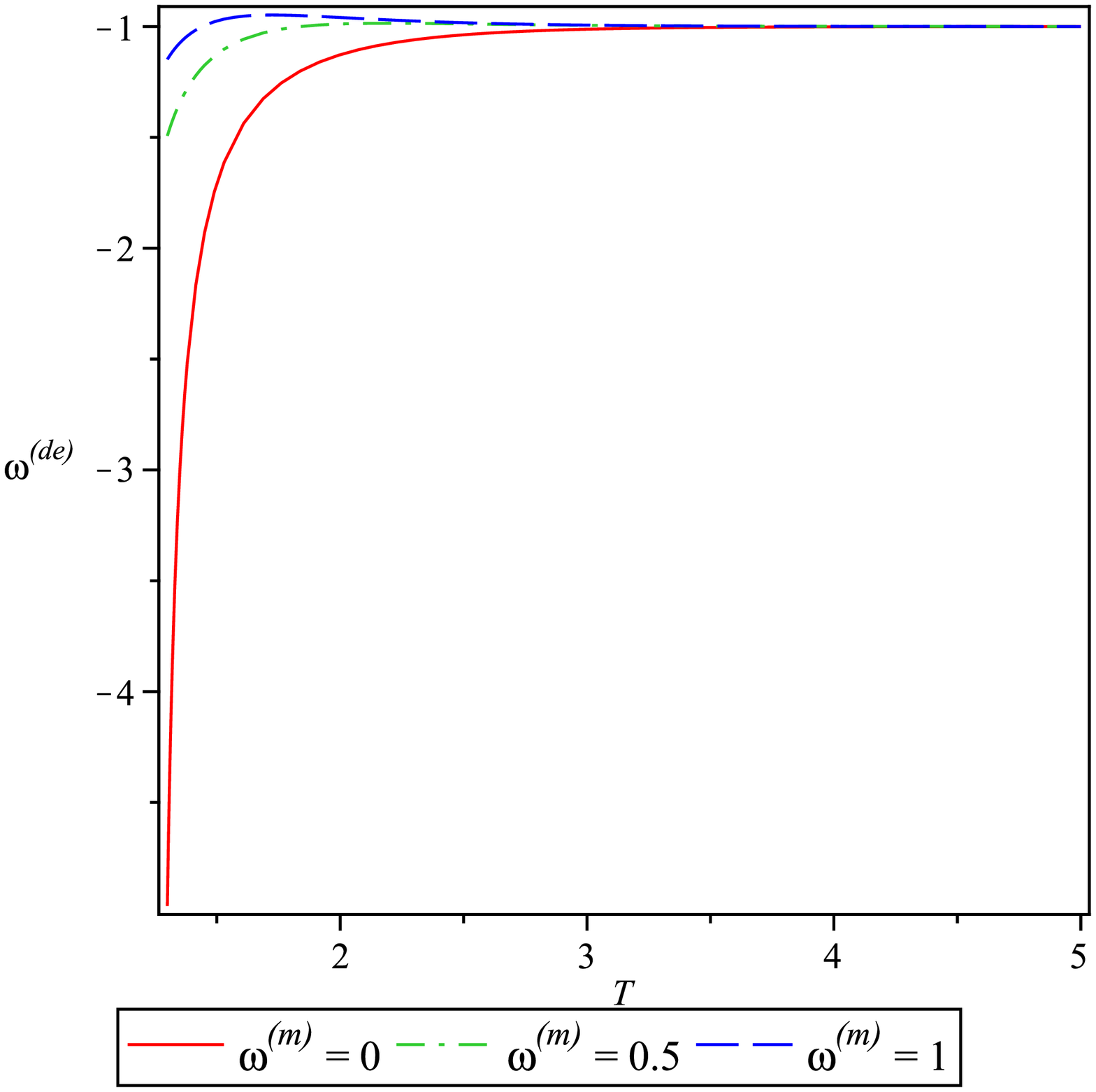}
\caption{The EoS DE parameter $\omega^{(de)}$ versus $T$ for $n = \beta = \alpha = \ell_{3} = l = 1$,
$\rho_{0} = 10$, $\kappa = 0.03$}
\label{fig:figure5}
\end{minipage}
\hspace{0.5cm}
\begin{minipage}[b]{0.5\linewidth}
\centering
\includegraphics[width=\textwidth]{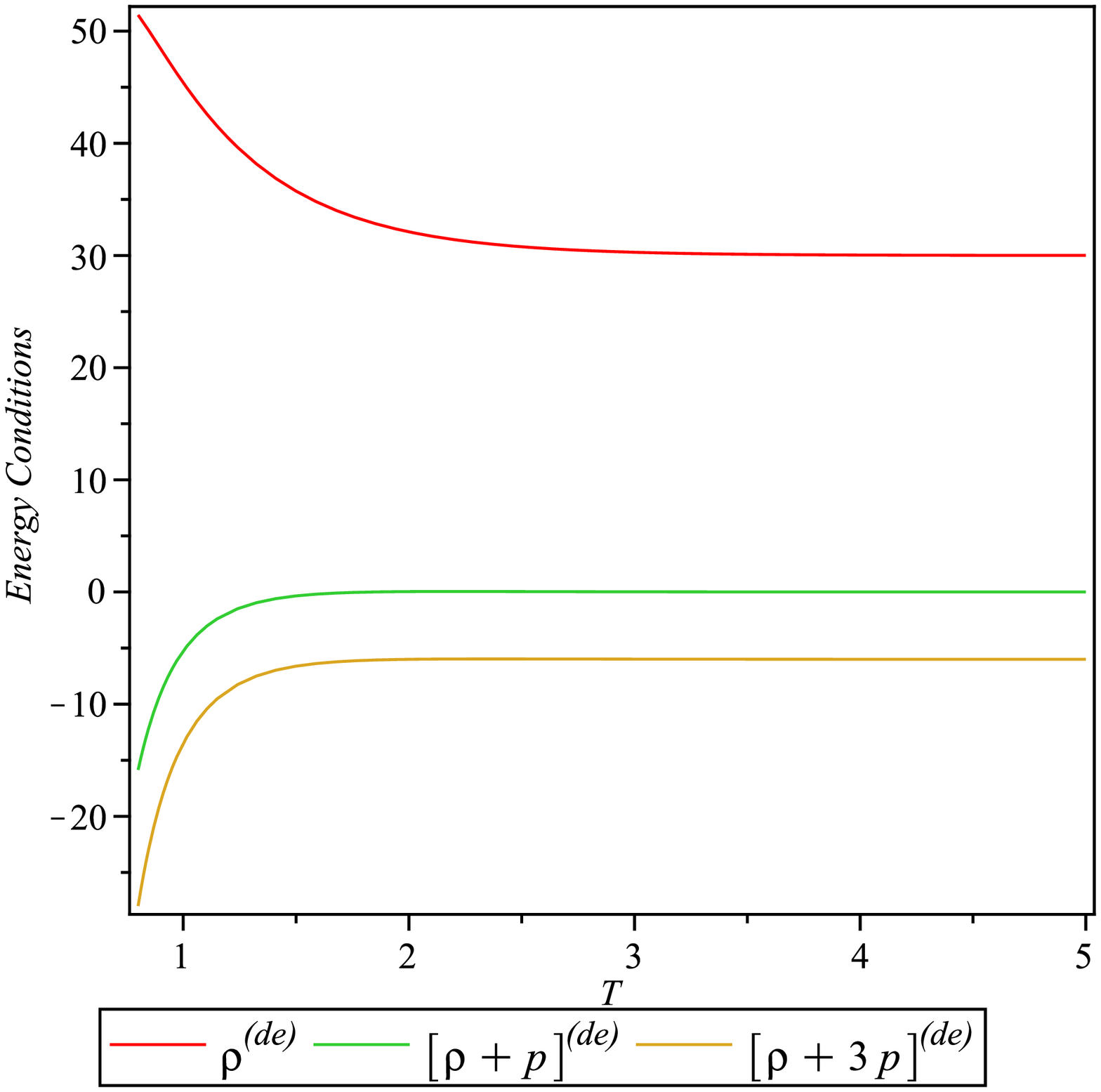}
\caption{The EoS DE parameter $\omega^{(de)}$ versus $T$ for $n = \beta = \alpha = \ell_{3} = l = 1$,
$\rho_{0} = 10, \omega_{m} = 0.5$, $\kappa = 0.03$}
\label{fig:figure6}
\end{minipage}
\end{figure}
\section{Interacting Two-Fluid Model}
In this section we consider the interaction between dark and barotropic fluids. For this purpose we can write
the continuity equations for barotropic and dark fluids as
\begin{equation}
\label{eq56}\dot{\rho}^{(m)} + 3\frac{\dot{a}}{a}\left(\rho^{(m)} + p^{(m)}\right) = \dot{\rho}^{(m)} +
(1 + \omega^{(m)})\rho^{(m)}(2n + 1)\frac{\dot{C}}{C} = Q,
\end{equation}
and
\begin{equation}
\label{eq57}\dot{\rho}^{(de)} + 3\frac{\dot{a}}{a}\left(\rho^{(de)} + p^{(de)}\right) = \dot{\rho}^{(de)} +
(1 + \omega^{(de)})\rho^{(de)}(2n + 1)\frac{\dot{C}}{C} = - Q.
\end{equation}
The quantity $Q$ expresses the interaction between the dark components. Since we are interested in
 an energy transfer from the dark energy to dark matter, we consider $Q > 0$. $Q > 0$, ensures that the second
law of thermodynamics is fulfilled \cite{ref118}. Here we emphasize that the continuity Eqs. (\ref{eq56}) and
(\ref{eq57}) imply that the interaction term ($Q$) should be  proportional to a quantity with units of inverse of
time i.e $Q\propto \frac{1}{t}$. Therefor, a first and natural candidate can be the Hubble factor $H$ multiplied
with the energy density. Generally, $Q$ could be taken in any of the following forms: (i) $Q\propto H\rho^{(m)}$ \cite{ref119, ref120}, (ii)
$Q\propto H\rho^{(de)} \cite{ref118, ref121}$, or (iii) $Q\propto H(\rho^{(de)}+\rho^{(m)})$ \cite{ref122}-\cite{ref124}.
It is worth to mention that because of our lack of knowledge of the nature of DE and DM, one can choose any of the above specific form of the interaction term $Q$. Moreover, as noted in ref \cite{ref85}, describing the interaction between the dark components of the universe through a microphysical model is not available today. In our study, Following Amendola et al. \cite{ref119} and Gou et al. \cite{ref120}, we consider
\begin{equation}
\label{eq58}Q = H \kappa \rho^{(m)},
\end{equation}
where $\kappa$ is a coupling coefficient which can be considered as a constant or variable parameter of redshif. As it is shown by Guo et al. \cite{ref115}, the combination of the SNLS, CMB, and BAO databases marginalized over a present dark energy density gives stringent constraints on the coupling, $-0.08 < \kappa < 0.03$ ($95\%$ C.L.) in the constant coupling model and $-0.4 < \kappa_{0} <0.1$ ($95\%$ C.L.) in the varying coupling model, where $\kappa_{0}$ is a present value.\\

Using Eq. (\ref{eq58}) in Eq. (\ref{eq56}) and after integrating, we obtain
\begin{equation}
\label{eq59}\rho^{(m)} = \rho_{0}C^{-(2n + 1)(1 + \omega^{(m)} - \kappa)} = \rho_{0} \l \sinh^{-3(1 + \omega^{(m)} -\kappa)}
{(\beta T)},
\end{equation}
where $\l = \ell_{3}^{-(2n + 1)(1 + \omega^{(m)} -\kappa)}$. Note that here for simplicity a constant coupling constant, $\kappa$, is considered.\\

By using Eqs. (\ref{eq19}), (\ref{eq20}) and (\ref{eq55}) in Eqs. (\ref{eq15}) and (\ref{eq12}), we obtain
\begin{equation}
\label{eq60} \rho^{(de)} = n(n + 2)\frac{\dot{C}^{2}}{C^{2}} - \frac{\alpha^{2}}{C^{2n}} - \rho_{0}C^{-(2n + 1)
(1 + \omega^{(m)} - \kappa)},
\end{equation}
and
\begin{equation}
\label{eq61} p^{(de)} = - \left[2n\frac{\ddot{C}}{C} + n(3n - 2)\frac{\dot{C}^{2}}{C^{2}} - \frac{\alpha^{2}}{C^{2n}}
\right] - \rho_{0}(\omega^{(m)} - \kappa)C^{-(2n + 1)(1 + \omega^{(m)} - \kappa)} .
\end{equation}
Using Eq. (\ref{eq26}) in Eqs. (\ref{eq60}) and (\ref{eq61}), we obtain the values of $\rho^{(de)}$
and $p^{(de)}$ as
\begin{equation}
\label{eq62}
\rho^{(de)} = \frac{9n(n + 2)\beta^{2}}{(2n + 1)^{2}}\coth^{2}(\beta T) - \alpha^{2}\ell_{3}^{-2n}
\sinh^{-\frac{6n}{2n + 1}}{(\beta T)} -\rho_{0} \l \sinh^{-3(1 + \omega^{(m)} -\kappa)}{(\beta T)},
\end{equation}
and
\[
 p^{(de)} = - \left[\frac{6n\beta^{2}}{(2n + 1)} + \frac{3n(5n - 2)\beta^{2}}{(2n + 1)^{2}}\coth^{2}{(\beta T)} -
\alpha^{2}\ell_{3}^{-2n}\sinh^{-\frac{6n}{2n + 1}}{(\beta T)}\right]
\]
\begin{equation}
\label{eq63}
 -\rho_{0}(\omega^{(m)} - \kappa)\l \sinh^{-3(1 + \omega^{(m)} - \kappa)}{(\beta T)}
\end{equation}
respectively.  \\

Also the EoS parameter for DE ($\omega^{(de)}$) is obtained as
\begin{equation}
\label{eq64}\omega^{(de)} = - \left[\frac{\frac{6n\beta^{2}}{(2n + 1)} + \frac{3n(5n - 2)\beta^{2}}{(2n + 1)^{2}}
\coth^{2}{(\beta T)} - \alpha^{2}\ell_{3}^{-2n}\sinh^{-\frac{6n}{2n + 1}}{(\beta T)} + \rho_{0}(\omega^{(m)} -
\kappa)\l \sinh^{-3(1 + \omega^{(m)} - \kappa)}{(\beta T)}}{\frac{9n(n + 2)\beta^{2}}{(2n + 1)^{2}}\coth^{2}(\beta T)
- \alpha^{2}\ell_{3}^{-2n}\sinh^{-\frac{6n}{2n + 1}}{(\beta T)} -\rho_{0} \l \sinh^{-3(1 + \omega^{(m)} -\kappa)}
{(\beta T)}}\right].
\end{equation}
The behavior of EoS ($\omega^{(de)}$) for dark energy in term of cosmic time $T$ is shown in Fig. $5$. It is observed
that the EoS parameter is an increasing function of time and the rapidity of its increase at the early stage depends on
the value of $\omega^{(m)}$, while later on it tends to the same constant value (i.e, $-1$) independent to it.\\

The expressions for the matter-energy density $\Omega^{(m)}$, dark-energy density $\Omega^{(de)}$ and the density
parameter $\Omega$ are same as given by Eqs. (\ref{eq40})$-$(\ref{eq42}) in the non-interacting case. \\

Based on Eqs. (\ref{eq62}) and (\ref{eq63}), the left hand side of energy conditions have been plotted in Figure $6$ for
different values of $T$. From Figure $6$, for $\omega^{(m)}\leq 1$ (i.e. phantom model) (also see Figure $5$) , we
observe that
$$
(i) ~ ~ \rho^{(de)} \geq 0,  ~ ~ ~ (ii) ~ ~  \rho^{(de)} + p^{(de)} \leq 0,  ~ ~ ~   (iii) ~ ~ \rho^{(de)} + 3p^{(de)} < 0.
$$
Thus, from above expressions, we observe the phantom model violates both the strong and weak energy conditions,
as expected. \\

\begin{figure}[ht]
\centering
\includegraphics[width=8cm,height=8cm,angle=0]{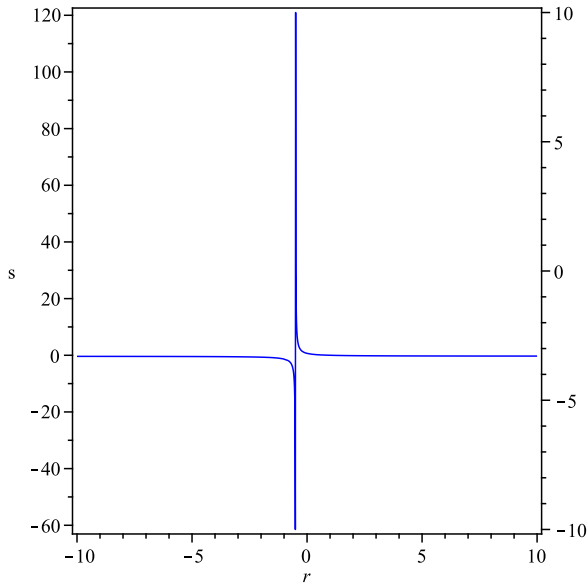} \\
\caption{The variation of $s$ against $r$}
\end{figure}

Studying the interaction between the dark energy and ordinary matter will open a possibility of detecting the dark
energy. It should be pointed out that evidence was recently provided by the Abell Cluster A586 in support of the
interaction between dark energy and dark matter \cite{ref125, ref126}. We observe that in non-interacting case both
open and flat universes can cross the phantom region whereas in interacting case only open universe can cross
phantom region. \\

Sahni et al. \cite{ref127}, Alam et al. \cite{ref27} have introduced a pair of parameters $\{r,s\}$, called
Statefinder parameters. In fact, trajectories in the $\{r,s\}$ plane corresponding to different cosmological
models demonstrate qualitatively different behaviour. The Statefinder parameters can effectively differentiate
between different form of dark energy and provide simple diagnosis regarding whether a particular model fits
into the basic observational data. The above Statefinder diagnostic pair has the following form:

\begin{equation}
\label{eq65} r = 1 + 3\frac{\dot{H}}{H^{2}} + \frac{\ddot{H}}{H^{3}} \; \; \mbox{and} \; \;
s = \frac{r - 1}{3(q -\frac{1}{2})} \; .
\end{equation}
For our model, the parameters $\{r,s\}$ can be explicitly written in terms of $T$ as
\begin{equation}
\label{eq66} r =  \tanh^{2}(\alpha T) \; ,  \; \; s = \frac{1}{3} \left[\frac{sech^{2}(\alpha T)}
{\tanh^{2}(\alpha T) + \frac{1}{2}}\right] \; .
\end{equation}
So the relation between $r$ and $s$ has the implicit form:
\begin{equation}
\label{eq67} 3(2r + 1)s + 2(r - 1) = 0.
\end{equation}
From Fig. $7$, we observe that $s$ is negative when $r \geq 1$. The figure shows that the universe starts from an
asymptotic Einstein static era ($r \to \infty, s \to - \infty$) and goes to the $\Lambda$CDM model ($r = 1, s = 0$).
\section{Concluding Remarks}
In this paper, we have studied a spatially homogeneous and anisotropic Bianchi type-III space time filled
with barotropic fluid and dark energy possessing dynamic energy density. The role of two-fluid either
minimally or directly coupled in the evolution of the dark energy parameter has been investigated by
considering the scalar expansion in the model is proportional to the shear scalar. The field equations
have been solved exactly with suitable physical assumptions. To prevail the deterministic solution we
choose the scale factor $a(t) = \sinh(\beta T)$, which yields a time-dependent deceleration parameter,
representing an accelerating model. The solutions satisfy the energy conservation Eq. (\ref{eq17}) and
the Raychaudhuri Eq. (\ref{eq18}) identically. Therefore, exact and physically viable Bianchi type-III
model has been obtained. The main features of the model are as follows: \\

$\bullet$ For $n = 1$ the anisotropic parameter $A_{m}$ tends to zero. Hence, the present model is
isotropic at $n = 1$.\\

$\bullet$ The derived DE model represents an acceleration universe (see, Figure $4$) which is in good
agreement with recent observations \cite{ref1}$-$\cite{ref3}. \\

$\bullet$ In non-interacting two-fluid model, we observe that for $\omega^{(m)} \geq 1$,  $\omega^{(de)}$
varies from non-dark region crossing to quintessence region and ultimately approaches to cosmological
constant region ($\omega^{(de)} = -1$). But for $\omega^{(m)} < 1$, the variation of $\omega^{(de)}$ starts
from super phantom region crossing phantom region ($\omega^{(de)} < -1$) and finally approaches to
cosmological constant region ($\omega^{(de)} = -1$) (see Fig. $1$). \\

$\bullet$ In interacting two-fluid model, we observe that $\omega^{(de)}$ is rapidly increasing function of
time and ultimately approaches to cosmological constant region ($\omega^{(de)} = -1$) (see Fig. 5). \\

$\bullet$ In non-interacting and interacting two-fluid models, we obtained that phantom  model violates both
the strong  and weak energy conditions, as expected (see, Figures $3$ \& $6$). \\

$\bullet$ In both non-interacting and interacting two-fluid scenario, the total density parameter ($\Omega$)
approaches to $1$ for sufficiently large time (see, Figure $2$) which is reproducible with current observations. \\

$\bullet$ The cosmic jerk parameter in our descended model is also found to be in good agreement with
the recent data of astrophysical observations namely the gold sample of type Ia supernovae \cite{ref96},
the SNIa data from the SNLS project \cite{ref97}, and the X-ray galaxy cluster distance measurements
\cite{ref118}. \\

$\bullet$ In both non-interacting and interacting two-fluid scenario, it is observed that such DE models are also
in good harmony with current observations. Thus, the solutions demonstrated in this paper may be useful for better
understanding of the characteristic of anisotropic DE in the evolution of the universe within the framework of
Bianchi type-III space-time. \\

$\bullet$ The Statefinder pair $\{r,s\}$ enable the behaviour of different stages of the evolution of the universe
i.e. the universe starts from asymptotic Einstein static era ($r \to \infty, s\to - \infty$) and goes to $\Lambda$
CDM model ($r = 1, s = 0$). \\

Finally, the solutions presented in this work can be one of the potential candidates to describe the observed
universe.

\section*{Acknowledgments}
The authors (A. Pradhan \& R. Jaiswal) would like to thank the Inter-University Centre for Astronomy and Astrophysics
(IUCAA), Pune, India for providing facility \& support where part of this work was done. The authors (A. Pradhan \& R.
Jaiswal) also gratefully acknowledge the financial support (Project No. C.S.T./D-1536) in part by State Council of
Science and Technology, Uttar Pradesh, India.

\end{document}